\documentclass[twoside,11pt]{entics} 

\usepackage{enticsmacro}

\usepackage{proof}
\usepackage{FMC}

\newcommand\coloneqq{\mathrel{{:}{:}{=}}}

\newcommand\JUMP{\mathcal J}

\newcommand\VAR{\mathcal V}

\newcommand\cbv{\textsf{cbv}}
\newcommand\cbn{\textsf{cbn}}
\newcommand\cbpv{\textsf{cbpv}}

\newcommand\tm{\mathsf{t}}
\newcommand\val{\mathsf{v}}

\newcommand\vc[1]{\vcenter{\hbox{#1}}}

\def\conf{MFPS 2024}
\volume{NN}

\def\lastname{Heijltjes}

\begin{document}

\begin{frontmatter}

\title{The Functional Machine Calculus III: Choice\\\normalsize{Early Announcement}}

\author{Willem Heijltjes\thanksref{e}}
\address{Department of Computer Science\\University of Bath\\Bath, United Kingdom}  							
\thanks[e]{Email: \href{mailto:w.b.heijltjes@bath.ac.uk} {\texttt{\normalshape mailto:w.b.heijltjes@bath.ac.uk}}} 


\begin{abstract}
The Functional Machine Calculus (Heijltjes 2022) is an extension of the lambda-calculus that preserves confluent reduction and typed termination, while enabling both call-by-name and call-by-value reduction behaviour and encoding the computational effects of mutable higher-order store, input/output, and probabilistic computation. In this note the calculus is extended to capture exception handling and loop constructs.
\end{abstract}


\end{frontmatter}


\section{Introduction}

The Functional Machine Calculus (FMC)~\cite{Heijltjes-2022,Barrett-Heijltjes-McCusker-2023} is a new approach to combining the $\lambda$-calculus---as the foundation of functional programming---with computational effects. It takes a view of the $\lambda$-calculus as an instruction language for an abstract machine with a single stack in the style of Krivine~\cite{Krivine-2007}, where \emph{application} is \emph{push}, \emph{abstraction} is \emph{pop}, and \emph{variable} is \emph{execute}. To accommodate effects, the calculus introduces the following two extensions~\cite{Heijltjes-2022}.

\begin{description}
	\item[Locations] 
Multiple stacks on the machine, each named by a \emph{location}, allow the encoding of various effects via push and pop actions: \emph{mutable higher-order store}, as stacks of depth at most one; \emph{input/output}, as pop-only respectively push-only streams; and \emph{probabilities} and \emph{non-determinism} as probabilistically respectively non-deterministically generated streams.
	
	\item[Sequencing] 
The introduction of \emph{sequential composition} and its unit, imperative \emph{skip}, gives control over evaluation behaviour away from strict call--by--name, and allows the encoding of Plotkin's call--by--value $\lambda$-calculus~\cite{Plotkin-1975}, Moggi's computational metalanguage~\cite{Moggi-1991}, and Levy's call--by--push--value~\cite{Levy-2003}.
\end{description}

Encoding effects into the generalized operators of the calculus, rather than introducing primitives, means that two key properties of the $\lambda$-calculus are preserved.

\begin{description}
	\item[Confluence]
Reduction in the FMC is confluent in the presence of effects. This is a consequence of the separation of operational behaviour, which governs the machine, from local reduction behaviour, which is the interaction of consecutive push and pop actions. Reduction equivalence for state then implements the algebraic laws of Plotkin and Power~\cite{Plotkin-Power-2002}.

	\item[Types]
The FMC can be simply typed, which conveys strong normalization and termination of the machine. This gives a solution to the problem of typing higher-order store: \emph{Landin's Knot}~\cite{Landin-1964}, which encodes recursion via higher-order store, cannot be typed (in its full generality).
\end{description}

This paper introduces a third extension to the FMC, \emph{choice}, to include a wider range of computational behaviours: \emph{constants}, \emph{conditionals}, \emph{data constructors}, \emph{exception handling}, and \emph{loops}. These have in common that, semantically, they are modelled by \emph{sums} or \emph{coproducts}: for example, the Booleans are given by the type $1+1$, the error monad is given by the functor $TX=E+X$ for a set of exceptions $E$, and loops are modelled by taking a map in $A\to A+B$ to one in $A\to B$ (looping on $A$, exiting on $B$)~\cite{Bloom-Esik-1993}. Together, these will be referred to as \emph{choice} constructs.

The aim is sixfold. First and second, to preserve \emph{confluence} and \emph{types}: the resulting calculus should support a natural, confluent reduction relation, and a notion of simple types that guarantees termination of the machine and strong normalization of reduction (in the absence of loops). Third, \emph{minimality}: choice constructs should be captured with as few syntactic operators as possible, avoiding any overlap in functionality and minimizing the interactions or reductions governing the semantics of the calculus. Fourth, \emph{operational semantics}: the calculus should continue to be an instruction language for a simple and natural abstract machine. Fifth, \emph{seamless integration}: different effects should combine seamlessly, without requiring lifting operations. Finally, the FMC has a natural \emph{first-order restriction}, where function arguments are restricted to be (first-order) values, not arbitrary terms. The sixth aim is to preserve this restriction, which ensures that choice constructs are independent of the calculus being first-order or higher-order.

The approach has been to reconsider the notion of \emph{choice} from first principles, with the aim of capturing coproducts and the constructs that they model in a simple and natural way, satisfying the six criteria above. This led to three (mostly) standard syntactic constructions, which however interact with the stack in subtle ways to give new and unexpected reduction behaviours. The resulting calculus is in some ways highly familiar, yet simultaneously in other ways novel and surprising.

In contrast with stateful effects, confluence and type safety are expected for exception handling. The main results are to integrate exceptions seamlessly with stateful effects, to support natural operational and denotational semantics, and to capture a wide range of behaviours with an elegant, minimal syntax. This note will discuss the background literature, introduce the selected \emph{choice} constructs from operational considerations, formally define the calculus and its type system, and demonstrate how it captures existing formulations. For simplicity of exposition, the calculus will omit the \emph{locations} modification, and feature only \emph{sequencing} and \emph{choice}. Proofs are incomplete at the time of writing, and hence omitted, leaving the intended theorems as conjectures.


\section{Background}

The study of effects in the $\lambda$-calculus is broad and varied. The overarching problem is that effects are \emph{sequential} when modelled through global updates, while the $\lambda$-calculus is fundamentally \emph{denotational}, governed by an equational theory. The contradiction manifests concretely as the loss of \emph{confluence}, so that the choice of reduction strategy becomes salient, and syntactic control over evaluation behaviour becomes necessary. Semantic reasoning about programs, a main strength of the $\lambda$-calculus, is severely impaired---and much research has the objective of restoring it. To quote Filinski in 2011:

\begin{quote}
Yet few would confidently claim that programs with computational effects are by now as well understood, and as thoroughly supported by formal reasoning techniques, as types and terms in purely functional settings. \hfill--- Filinski~\cite{Filinski-2011}
\end{quote}

Since Landin's seminal work~\cite{Landin-1964,Landin-1965,Landin-1966} most practical functional programming languages employ a call--by--value (\cbv) semantics, as formalized by Plotkin~\cite{Plotkin-1975}, where call--by--name (\cbn) behaviour is forced locally with \emph{thunks}. However, for reasoning about programs a \cbn-semantics is desirable. The inefficiency of \cbn\ may be overcome by \emph{graph reduction}~\cite{Wadsworth-1971} (or \emph{lazy evaluation}, or \emph{call--by--need}~\cite{Turner-1985,PeytonJones-1987,Ariola-Felleisen-Maraist-Odersky-Wadler-1995}), but the imposed notion of sequentiality is opaque, and a semantic approach to effects becomes more urgent.

Among various proposals, such as taming destructive updates through \emph{interference control}~\cite{Reynolds-1978,OHearn-Power-Takeyama-Tennent-1999} or \emph{uniqueness types} \cite{Smetsers-Barendsen-vanEekelen-Plasmeijer-1993}, and explictly typing effects in \emph{type and effect} systems~\cite{Gifford-Lucassen-1986,Jouvelot-Gifford-1991,Talpin-Jouvelot-1994,Nielson-Nielson-1999}, Moggi's account of effects as monads~\cite{Moggi-1991} became prominent. It casts sequentiality as composition in the Kleisli category of a monad~\cite{MacLane}, which may then be presented as sequential programming through syntactic sugar, such as Haskell's \emph{do}-notation. 


However, monads do not compose, raising the issue of how to combine multiple effects. Haskell 98 resorts to cumbersome stacks of \emph{monad transformers} \cite{Liang-Hudak-Jones-1995} (their limitations are discussed in \cite{Kiselyov-Sabry-Swords-2013}). One solution is to take sums of monads~\cite{Luth-Ghani-2002}; another, added in Haskell 2010, is to weaken monads to \emph{applicative} (or \emph{monoidal}) functors \cite{McBride-Paterson-2008}, which do compose. Both however need intrusive \emph{lifting} functions to access different effects, suggesting that the problem is not merely with compositionality, but instead with explicit composition as such.

Levy's~\emph{call--by--push--value} (\cbpv)~\cite{Levy-2003} unifies effects in a single monad (not unlike the IO monad in Haskell~\cite{PeytonJones-Wadler-1993}) formed by an adjunction separating \emph{values} and \emph{computations}. It arose in the context of two closely related alternatives. Power, Robinson, and Thielecke sought to connect Hasegawa's $\kappa$-calculus~\cite{Hasegawa-1995} with a semantics in \emph{premonoidal} categories~\cite{Power-Robinson-1997,Power-Thielecke-1999}, which impose sequentiality on composition in a monoidal category by weakening its naturality laws---a semantics later applied to \cbpv~\cite{Levy-Power-Thielecke-2003}. This was mirrored by Hughes' Arrows for Haskell~\cite{Hughes-2000,Lindley-Wadler-Yallop-2010}, which generalise Kleisli composition of monads in a similar way~\cite{Atkey-2011}. The similarities between \cbpv\ and $\kappa$-calculus are striking: both feature sequentiality through an operational semantics in a stack machine, a decomposition of computations and values, and an embedding of \cbn- and \cbv-$\lambda$-calculus. It was \cbpv\ that gained traction, being extended~\cite{Egger-Mogelberg-Simpson-2014} and rediscovered~\cite{Ehrhard-Guerrieri-2016} through linear logic and forming the basis of \emph{effect handlers}~\cite{Plotkin-Pretnar-2013}.

Effect handlers~
\cite{Plotkin-Pretnar-2013,Kammar-Lindley-Oury-2013,Wu-Schrijvers-2015,Marsik-Amblard-deGroote-2021} combine $\cbpv$ with \emph{algebraic effects}~\cite{Plotkin-Power-2002,Hyland-Plotkin-Power-2006,Hyland-Levy-Plotkin-Power-2007}, a view of effects as algebraic theories that enables equational reasoning with effect operators, though not reduction. This is implemented by restricting \emph{continuations} to enforce \emph{sequentiality} (a restriction that is sometimes relaxed), generalising \emph{exception handlers}~\cite{Benton-Kennedy-2001}.

Continuations as a means of controlling sequential behaviour date back to ALGOL60 and Landin's J-operator~\cite{Landin-1965} used to model its \emph{go to} construct (Reynolds provides an overview of this early history~\cite{Reynolds-1993}), subsequently developed in the context of the \cbv-$\lambda$-calculus~\cite{Felleisen-Friedman-1986,Felleisen-Friedman-Kohlbecker-Duba-1987}. Three developments then gave continuations a central place in programming language theory and practice: the discovery of a Curry--Howard relation with classical logic~\cite{Griffin-1990}, soon after expressed in Parigot's $\lambda\mu$-calculus~\cite{Parigot-1992,Streicher-Reus-1998}; their usefulness in compilers~\cite{Appel-1992}; and the charactarization of the continuation monad as universal among effect monads~\cite{Filinski-1994,Filinski-1996}. 

The distinction between continuations and \emph{exception handling} is wafer-thin in a higher-order language, and exceptions in $\lambda$-calculus are frequently studied through continuations~\cite{DeGroote-1995,Ong-Stewart-1997,Bierman-1998,Crolard-1999,Kameyama-Sato-2002,Lebresne-2009,VanBakel-2019}. Subtle differences in reduction arise depending on the choice of semantics: in classical logic and $\lambda\mu$-calculus, or in intuitionistic coproducts as the \emph{exceptions-as-values} approach generally models them~\cite{Wadler-1985,Spivey-1990,PeytonJones-Reid-Henderson-Hoare-Marlow-1999}, or in intuitionistic continuations as do the exception handlers~\cite{Benton-Kennedy-2001,Levy-2006-Exceptions} that inspired effect handlers~\cite{Plotkin-Pretnar-2013}.

The Functional Machine Calculus itself~\cite{Heijltjes-2022,Barrett-Heijltjes-McCusker-2023} originates in a confluent probabilistic $\lambda$-calculus~\cite{DalLago-Guerrieri-Heijltjes-2020}, and while developed independently, many individual aspects are familiar. \emph{Sequencing}, which locates sequentiality in the operational semantics of a stack machine, appeared as $\kappa$-calculus~\cite{Hasegawa-1995,Power-Thielecke-1999}, in compiler theory~\cite{Douence-Fradet-1998}, and in practice as \emph{concatenative programming}~\cite{Herzberg-Reichert-2009,Pestov-Ehrenberg-Groff-2010,Mihelic-Steingargner-Novitzka-2021}. \emph{Locations} are akin to the \emph{channels} of process calculi~\cite{Hoare-1978,Milner-1980,Milner-Parrow-Walker-1992}, and their encoding of mutable strore is that of Haskell's \emph{MVar} mutable variables for shared-memory concurrency~\cite{PeytonJones-Gordon-Finne-1996}. A close overall relative are \emph{action calculi}~\cite{Mifsud-Milner-Power-1995,Milner-1996}. The FMC as a whole is nevertheless novel and surprising, as the introduction sets out.

The \emph{choice} extension in this note uses standard syntactic constructions for exception handling, with subtly novel reduction behaviour due to the interaction with the stack. It takes a semantics in intuitionistic sums, and avoids swerving into continuations-territory by the design requirements of \emph{simple types} and \emph{first-order restriction}. Key inspirations are compiler optimizations through sequent calculus and sequentially-used continuations~\cite{Downen-Maurer-Ariola-PeytonJones-2016,Maurer-Downen-Ariola-PeytonJones-2017}, from which the terminology of \emph{jumps} and \emph{joins} is adopted. The \emph{loop} construct is believed to be a new syntactic interpretation of the standard semantics for iteration~\cite{Bloom-Esik-1993}, dual to recursion~\cite{Simpson-Plotkin-2000}.


\section{Development}

The FMC extends the $\lambda$-calculus with \emph{sequential composition} $(\term{M;N})$ and \emph{skip} $(\term*)$, its unit, the empty instruction sequence. In an imperative context, \emph{skip} signifies \emph{successful termination}. The FMC introduces this concept into the Krivine Machine (which traditionally halts with an abstraction and empty stack, or a variable), and it is a key feature underlying many aspects of the calculus, including the design of the type system, the semantics, and the termination and normalization proofs~\cite{Heijltjes-2022,Barrett-Heijltjes-McCusker-2023}.

The central idea of this paper is to generalise \emph{skip} to allow not just \emph{success}, but arbitrary notions of \emph{failure}, i.e.\ exceptions or exit codes. This is modeled by a set of \emph{jumps} $\JUMP$, with \emph{skip} $(\term*\in\JUMP)$ the default value. As a term, a jump $\term j$ corresponds to raising an exception. The machine may terminate with any jump $\term j\in\JUMP$ as exit status, plus a stack of return values. Semantically, a term returns a coproduct over stacks, indexed in a finite set $J\subset\JUMP$.

Standard sequencing $\term{M;N}$ evaluates $\term M$ to $\term *$, and then continues with $\term N$, expressed in the reduction rule $\term{*;N}\rw\term{N}$. When $\term M$ may evaluate to any jump $j$, sequencing is likewise generalized by making it conditional on a particular jump, to a \emph{join} $\term{M;j->N}$: if $\term M$ terminates in $j$ then continue as $N$ (and otherwise discard $N$). This makes $\term{M;j->N}$ a standard \emph{try/catch} construct, $\mathsf{try}~\term M~\mathsf{catch}~\term j~\term N$. The behaviour is formalized by the reduction rules $\term{j;j->N}\rw\term N$ and $\term{i;j->N}\rw\term i$ (where $i\neq j$). The original sequencing $\term{M;N}$ becomes an abbreviation for the default case $\term{M;*->N}$.

The construction lets us cast constants as jumps, for example the Booleans as $\term\top$ and $\term\bot$ with the conditional $\mathsf{if}~\term B~\mathsf{then}~\term{M}~\mathsf{else}~\term{N}$ given by two joins as $\term{B;\top->M;\bot->N}$ (which associates as $\term{(B;\top->M);\bot->N}$, and assumes that both $\term M$ and $\term N$ exit with \emph{skip}). Our \emph{join} is then also a standard \emph{case switch}:
\[
	\mathsf{switch}~\term M~(\mathsf{case}~\term{j_1}:~\term{N_1} \dots \mathsf{case}~\term{j_n}:~\term{N_n})
\]
becomes $\term{M;j_1->N_1 \dots ;j_n->N_n}$. Such switches may have different \emph{fall-through} policies (whether after a matching case also subsequent cases are considered), often with a \emph{break} or \emph{return} keyword to exit; it will come as no surprise that these are readily modelled with jumps and joins, where \emph{break} is another jump.

A subtle but key aspect is the difference between returning a constant as a value on the stack, and evaluating it as a jump. A standard function returning a Boolean should be interpreted in the former way, pushing $\term\top$ or $\term\bot$ onto the stack and exiting with $\term*$, and indeed terms are generally expected to exit successfully. To \emph{evaluate} a constant, triggering subsequent join constructs, is the domain of conditionals, case switches, and exception handling.

Joins as conditional sequencing give a clean approach to \emph{iteration}. A \emph{loop} $\term{M^j}$ evaluates $\term M$, then loops on $\term j$ or exits on other jumps. This is implemented by the reduction $\term{M^j}\rw \term{M ; j -> M^j}$. A standard \emph{do--while} loop $(\mathsf{do}~\term M~\mathsf{while}~\term B)$ with a Boolean $B$ becomes $\term{(M;B)^\top;\bot->*}$, with the expected behaviour: it reduces to $\term{M ; B ; \top -> (M;B)^\top ; \bot -> *}$, which is the interpretation of
\[
	\term{M}~;~\mathsf{if}~\term B~\mathsf{then}~(\mathsf{do}~\term M~\mathsf{while}~\term B)~\mathsf{else}~\term{*}~.
\]
As before, control flow through \emph{break} or \emph{return} keywords may be modelled by viewing these as jumps, and positioning corresponding \emph{join} constructs outside the loop.

On an abstract machine, it is standard to manage sequencing with a \emph{continuation stack}, where $\term{M;N}$ pushes $\term N$, and \emph{skip} pops. The FMC, in the formulation used here, introduces this into the Krivine Machine. The generalization to jumps is captured on the machine by modifying the continuation stack to hold \emph{conditional terms}: $\term{M;j->N}$ pushes $\term{j->N}$, while $\term j$ inspects the head $\term{i->N}$ and pops $\term N$ if $i=j$, discarding it otherwise. A loop $\term{M^j}$ pushes $\term{j-> M^j}$ and continues as $\term M$. The three constructs thus represent natural instructions on the machine, with $j$ representing a forward jump and $\term{M^j}$ a controlled backward jump.

Types for the FMC indicate the input and output stacks of successful machine evaluation: a term of type $\type{s_n..s_1 => t_1..t_m}$ expects an input stack of terms typed $\type{s_1}$ through $\type{s_n}$ and produces a return stack with types $\type{t_1..t_m}$. Semantically, terms are morphisms in a category where objects are type vectors $\type{!t}=\type{t_1..t_m}$ (assuming a single stack for simplicity; with multiple stacks, objects are families of type vectors).
With jumps, return types become sums of jump-indexed type vectors, each written $\type{!t.j}$, rendering types as
\[
	\type{!s => (!t_1.j_1 + .. + !t_n.j_n)}~.
\]
For example, using $\type1$ for the empty stack the type for Booleans becomes $\type{1 => 1.\top + 1.\bot}$. As this illustrates, jumps (as terms) are the \emph{injections} of the coproduct. 

A join $\term{M;j->N}$ composes on the $j$-indexed summand; for example for a Boolean $\term B$ and $\term{M:1=>!t.*}$ we have $\term{B;\top->M:1=>!t.* + 1.\bot}$. Codiagonals (the natural maps $A+A\to A$) are given implicitly by two joins with the same return type \emph{and} jump. For example, if $\term{M}$ and $\term{N}$ are both typed $\type{1 => !t.*}$ then the conditional $\term{B;\top->M;\bot->N}$ will have the type $\type{1 => !t.*}$ as well. 

The types for \emph{loops} follow the standard semantics of iteration. Since $\term{M^j}$ iterates on $j$ and exits on other jumps, the return stack of $\term M$ for the jump $j$ must match its input stack, and this summand is removed for $\term{M^j}$. For example, if $\term{M : t => t.*}$ and $B$ is a Boolean, then $\term{M;B : t => t.\top + t.\bot}$, and $\term{(M;B)^\top : t => t.\bot}$; i.e.\ if this loop terminates, it will be with a return value of type $\type t$ and exit status $\bot$.

The three constructors \emph{jump} $\term j$, \emph{join} $\term{M;j->N}$, and \emph{loop} $\term{M^j}$ thus form a concise internal language for coproducts with iteration. Next, we consider these in the context of the FMC, as an extension to the $\lambda$-calculus viewed through the lens of the (simplified) Krivine machine.


\section{The FMC with Choice}

The FMC adopts a syntax that emphasises the machine perspective, writing \emph{application} $\term{M\,N}$ as $\term{[N].M}$ to mean \emph{push} $\term N$ and continue with $\term M$, and \emph{abstraction} $\term{lx.M}$ as $\term{<x>.M}$ for \emph{pop}, to continue with $\term M$ where the popped term replaces $\term x$. The full FMC~\cite{Heijltjes-2022} parameterizes the \emph{push} and \emph{pop} operations in a set of \emph{locations}~$\{a,b,c,\dots\}$, each indicating a stack on the machine, as $\term{[N]a.M}$ and $\term{a<x>.M}$. This extension is orthogonal to \emph{choice} as considered here, and for simplicity is omitted. The constructions are thus introduced into the plain $\lambda$-calculus, but they generalise \emph{sequencing} $\term{M;N}$ and \emph{skip} $\term*$ of the fragment of the FMC called the \emph{sequential $\lambda$-calculus}.

\begin{definition}
Let $x,y,z$ range over a countable set of \emph{variables} $\VAR$, and $i,j,k$ over a countable set of \emph{jumps} $\JUMP$ that includes the distinguished element $\term *$. \emph{Terms} are given by the following grammar.
\[
	\term{M,N,P,Q}
~\coloneqq~ \term x
~\mid~		\term{[N].M}
~\mid~		\term{<x>.M}
~\mid~		\term j
~\mid~		\term{N;j->M}
~\mid~		\term{M^j}
\]
\end{definition}

The constructors are: a \emph{variable} $\term x$, an \emph{application} or \emph{push action} $\term{[N].M}$, an \emph{abstraction} or \emph{pop action} $\term{<x>.M}$ which binds $\term x$ in $\term M$, a \emph{jump} $\term j$, a \emph{join} $\term{N;j->M}$, and a \emph{loop} $\term{M^j}$. The free variables of a term are written $\textsf{fv}(\term M)$, and capture-avoiding substitution of $\term N$ for $\term x$ in $\term M$ is written $\term{\{N/x\}M}$.

\begin{definition}
The \emph{reduction rules} are:
\[
\begin{array}{l@{\qquad}r@{}l@{\qquad}l}
		\textnormal{Beta}          & \term{[N].<x>.M}     &~\rw~\term{\{N/x\}M}
	\\	\textnormal{Select}        & \term{j;j->M}        &~\rw~\term{M}
	\\	\textnormal{Skip}          & \term{i;j->M}        &~\rw~\term{i}				& (i\neq j)
	\\	\textnormal{Unroll}        & \term{M^j}			  &~\rw~\term{M; j-> M^j}
	\\	\textnormal{Prefix (pop)}  &\term{(<x>.N);j->M}   &~\rw~\term{<x>.(N;j->M)}   & (x\notin\textsf{fv}(\term M))
	\\	\textnormal{Prefix (push)} & \term{([P].N);j->M}  &~\rw~\term{[P].(N;j->M)}
\end{array}		
\]
The \emph{reduction relation} $\rw$ is given by closing the above rules under any context. The reflexive--transitive closure of $\rw$ is written $\rws$.
\end{definition}

The \emph{beta}-reduction rule is that of the $\lambda$-calculus. The \emph{select} and \emph{skip} rules implement a jump from $\term j$ to the next join $\term{j -> M}$, skipping others in between. The two \emph{prefix} rules govern the interaction between the $\lambda$-calculus and sequencing. They implement the notion that \emph{push} and \emph{pop} actions are \emph{sequential} via the (standard) interaction of prefixing with concatenation.

\begin{proposition}
The normal forms $\term{N_0}$ of reduction $\rw$ are given by the following grammars.
\[
\begin{array}{l@{~\coloneqq~}r@{~\mid~}l@{}l}
		\term{N_0}  & \term{<x>.N_0}      & \term{N_1}
	\\	\term{N_1}  & \term{[N_0].N_1}    & \term{N_2} & ~\mid~\term{j}
	\\	\term{N_2}  & \term{N_2;j -> N_0} & \term{x}
\end{array}
\]
\end{proposition}

The machine will evaluate a term in the context of an \emph{argument stack} $S=\term{M_1}\dots\term{M_n}$, holding terms, and a \emph{continuation stack} $K=\term{(j_1->M_1})\dots(\term{j_n->M_n})$, holding conditional continuations $\term{j->M}$ to be used on a jump $\term j$ and discarded on any other jump.

\begin{definition}
The \emph{states} of the abstract machine are triples $(S,\term M,K)$ of an \emph{argument stack} $S$, a term $\term M$, and a \emph{continuation stack} $K$, where both stacks are defined as follows.
\[
S,T	~\coloneqq~ \e ~\mid~ S\,\term{M}
\qquad\qquad
K,L ~\coloneqq~ (\term{j->M})\,K~\mid~\e
\]
The \emph{transitions} of the machine are given by the following top-to-bottom rules.
\[
\begin{array}{c@{\qquad}cc}
	\step
	  {S}         {[N].M}{K} 
	  {S\,\term N}    {M}{K}
&	\step
	  {S}{N;j->M}               {K}
	  {S}{N}     {(\term{j->M})\,K}
\\ \\[-10pt]
	\step
	  {S\,\term N}   {<x>.M}{K}
	  {S}         {\{N/x\}M}{K} 
&	\step
      {S}{j}{(\term{j->M})\,K}
      {S}{M}               {K}
\\ \\[-10pt]
	\step
	  {S}{M^j}{K}
	  {S}{M}{(\term{j->M^j})\,K}
&	\step
      {S}{i}{(\term{j->M})\,K}
      {S}{i}               {K}	 & (i\neq j)
\end{array}
\]
A \emph{run} of the machine is a sequence of steps written as below left. A run is \emph{complete} if it terminates with a jump $\term j$ as term and an empty continuation stack, as below right.
\[
\steps SMK TNL \qquad\qquad \steps SMK Tj\e
\]
\end{definition}

The abstract machine defines the small-step operational semantics. A big-step semantics is given by composing complete runs, as follows.

\begin{definition}
The \emph{evaluation} relation $\evl SMTj$ is defined inductively as follows.
\[
\infer{\eval SjSj}{}
\qquad
\begin{array}{@{}c@{\qquad}c@{\qquad}c@{}}
	\infer{\eval S{[N].M}Tj}{\eval{S\,\term N}MTj}
&	\infer{\eval S{M;i->N}Tj}{\eval SMRi && \eval RNTj}
&	\infer[(i\neq j)]{\eval S{M;i->N}Tj}{\eval SMTj}
\\ \\[-10pt]
	\infer{\eval {S\,\term N}{<x>.M}Tj}{\eval S{\{N/x\}M}Tj}
&	\infer{\eval S{M^i}Tj}{\eval SMRi && \eval R{M^i}Tj}
&	\infer[(i\neq j)]{\eval S{M^i}Tj}{\eval SMTj}\quad
\end{array}
\]
\end{definition}

\begin{proposition}
Small-step and big-step operational semantics agree:
\[
	\steps SM\e Tj\e \quad\iff\quad \evl SMTj
\]
\end{proposition}

To relate reduction and machine evaluation, the reduction relation $\rw$ is extended to argument stacks by reducing on a single term: $T\,\term M\,\rw\,T\,\term N$ if $\term M\rw\term N$.

\begin{proposition}
Reduction commutes with machine evaluation: if $S\rws R$, $\term M\rws\term N$, and $\evl SMTj$ then there is a stack $U$ such that $T\rws U$ and $\evl RNUj$.
\end{proposition}

\begin{conjecture}
Reduction $\rw$ is confluent.
\end{conjecture}

\section{Types}

Types for the sequential $\lambda$-calculus are as follows: stacks are typed $S:\type{!s}$ by a vector of types, and terms are typed $\term{M: !s => !t}$ indicating the expected input and output stacks for a complete run of $\term M$ on the machine. To accommodate \emph{choice}, the return type is generalised to a sum of jump-indexed type vectors, each written $\type{!t.j}$, to give the following type structure: $\term{M: !s => (!t_1.j_1 + .. + !t_n.j_n)}$. 

\begin{definition}
\emph{Types} are given as follows, where $I,J\subset\JUMP$ denote a finite set of jumps.
\[
\begin{array}{l@{\qquad}l@{~\coloneqq~}l}
\textnormal{Types:}			& \type{r,s,t}	& \type{!s => !tJ}           \\
\textnormal{Stack types:}	& \type{!t}		& \type{t_1 .. t_n}           \\
\textnormal{Choice types:}	& \type{!tJ}    & \{\type{!t_j}\mid j\in J\}
\end{array}
\]
\end{definition}

Observe that the identity function on a stack of two elements is $\term{<x>.<y>.[y].[x].*}$, while $\term{<x>.<y>.[x].[y].*}$ is the symmetry. The order of input types follows that of the abstractions, so that the type for the above identity is $\type{st=>ts}$ where $\term{x:s}$ and $\term{y:t}$, and in general identity types are $\type{?t => !t}$ where $\type{?t}$ is the reverse vector to $\type{!t}$. Further notation includes: concatenation of type vectors by juxtaposition $\type{!s!t}$, and the empty vector as $\type{1}$. The disjoint union of choice types as families is written $\type{!sI+!tJ}$ where $I\cap J=\varnothing$, and $\type{!t.j}$ denotes the singleton family consisting of the vector $\type{!t}$ indexed by $j$. A choice type may then be written $\type{!t_1.j_1 + .. + !t_n.j_n}$ where the jumps are disjoint.

Semantically, types $\type{?s => !tJ}$ are function spaces $\type{!s}\rightarrow\type{!tJ}$, stack types $\type{t_1..t_n}$ are products $\type{t_1} \times \dots \times \type{t_n}$, and choice types $\{\type{!t_j}\mid j\in J\}$ are indexed sums $\sum_{j\in J}(\type{!t_j})$. \emph{Locations} are indexed products~\cite{Barrett-Heijltjes-McCusker-2023}, giving the duality between state and exceptions observed in~\cite{Dumas-Duval-Fousse-Reynaud-2012}.

\begin{definition}
A \emph{typing judgement} $\term{G |- M:t}$ assigns a term $\term M$ the type $\type t$ in the context $\Gamma$, where a context is a finite function from variables to types written as a comma-separated list $\term{x_1:t_1,..,x_n:t_n}$. A \emph{stack typing judgement} $\trm{G |- S:}\type{!t}$ assigns a stack the type vector $\type{!t}$. The \emph{typing rules} for the calculus are given in Figure~\ref{fig:types}.
\end{definition}

\begin{figure}
\[
\begin{array}{l@{\qquad}c}
	\textnormal{Variable:} &
	\vc{\infer{\term{G , x: t |- x: t}}{}}
\\ \\
	\textnormal{Application:} &
	\vc{
	\infer{\term{G |- [N].M : ?s => !tJ}}{
		\term{G |- N : r}
	  & \term{G |- M : r\,?s => !tJ}
	}}
\\ \\
	\textnormal{Abstraction:} &
	\vc{
	\infer
	  {\term{G |- <x>.M : r\,?s => !tJ}}
	  {\term{G , x:r |- M : ?s => !tJ}}
	}
\\ \\
	\textnormal{Jump:} &
	\vc{\infer{\term{G |- j : 1 => 1.j }}{}}
\\ \\
	\textnormal{Join:} &
	\vc{
	\infer{\term{G |- N;j->M : ?r => !tI}}{
		\term{G |- N: ?r => !tI + !s.j}
	  & \term{G |- M: ?s => !tI}
	}}
\\ \\
	\textnormal{Loop:} &
	\vc{\infer{\term{G |- M^j : ?s => !tI}}{\term{G |- M: ?s => !tI + !s.j}}}
\\ \\
	\textnormal{Stack expansion:} &
	\vc{\infer{\term{G |- M: ?r\,?s => (!s\,!t)J}}{\term{G |- M: ?r => !tJ}}}
\\ \\
	\textnormal{Sum expansion:} &
	\vc{\infer{\term{G |- M: ?r => !sI + !tJ}}{\term{G |- M: ?r => !tJ}}}
\\ \\
	\textnormal{Stack types:} &
	\vc{\infer{\term{G |- M_1 .. M_n : t_1..t_n}}{\{\term{G |- M_i : t_i}\}_{i\leq n}}}
\end{array}
\]
\caption{Typing rules}
\label{fig:types}
\end{figure}

\begin{proposition}
Subject reduction holds: if $\term{G |- M:t}$ and $\term M\rw\term N$ then $\term{G |- N:t}$.
\end{proposition}

\begin{conjecture}
Types guarantee machine termination: for a loop-free, closed, typed term $\term{|- M:!s=>!tJ}$ and stack $\trm{|- S:}\type{!s}$ there is a jump $\term j\in J$ and stack $\trm{|- T:}\type{!t_j}$ with $\type{!t_j}\in\type{!tJ}$ such that $\eval SMTj$.
\end{conjecture}

\begin{conjecture}
Types guarantee strong normalization: a loop-free, typed term $\term{G |- M:t}$ does not have an infinite reduction path in $\rw$.
\end{conjecture}


\section{Discussion}

The careful design of the choice constructions for the FMC has subtle implications, which manifest more clearly in the context of the stack machine, the operational semantics of the calculus. We consider these here.

First, we consider data constructors. Standard case constructs such as are found in ML or Haskell are a direct interpretation of the disjunction elimination rule of natural deduction,
\[
	\mathsf{case}~M~\mathsf{of}~(\mathsf{inl}~x~\to~N~;~\mathsf{inr}~y~\to~P)
\]
generalised to indexed sums (data types with constructors) and multiple parameters. As does ML~\cite{Appel-MacQueen-Miler-Tofte-1988}, the present approach unifies data constructors and exceptions, and renders the above as follows.
\[
	\term{M~;~{\mathsf{inl}} -> <x>.N~;~{\mathsf{inr}} -> <y>.P}
\]
Instead of parameterizing the constructors as $\mathsf{inl}~x$ and $\mathsf{inl}~y$, the two joins use a regular \emph{pop}, $\term{<x>.N}$ or $\term{<y>.P}$, to collect the return value from $\term M$. This is possible due to the nature of the FMC as a stack-based calculus, where $\term M$ leaves its return value on the stack: an application to a constructor $\mathsf{inl}\,Q$ translates direcly as $\term{[Q].{\mathsf{inl}}}$, pushing $\term Q$ to the stack and exiting with $\mathsf{inl}$. Construction and deconstruction generalize to any number of arguments, simplyfying the syntax from a \emph{sequence} of cases each with a \emph{sequence} of parameters, to the single constructor $\term{M;j->N}$. The correctness of the interpretation is guaranteed by typing, which ensures that $\term N$ does not exit with $\mathsf{inr}$ which would incorrectly trigger $\term{P}$ in the example above, or in the untyped case by strictly jumps into \emph{constants/constructors} and \emph{exceptions}, and maintaining that constants and constructors are returned on the stack.

The fall-through aspect of joins makes them a potential solution to the problem of \emph{extensible data types}~\cite{Swierstra-2008} (adding further constructors to an existing data type). However, the present exposition does not include \emph{inductive} types, or \emph{recursion} to compute with them, so at the moment this is only an interesting future direction.

Secondly, we will look at exception handling mechanisms. Though there are many variations, the underlying principle is the same: a computation may \emph{raise} or \emph{throw} an exception, which is then \emph{caught} by some \emph{handler}. First, we discuss the formulation by De Groote~\cite{DeGroote-1995}, for it raises an important issue. It considers exception handling in ML, and proceeds to give a variation on it via classical logic; we will refer to the original ML variant as \emph{coproduct-style}, and that of classical logic, \emph{continuation-style}. The syntax is as follows, slightly simplified and using $M,N$ for terms and $e$ for exceptions.
\[
	\mathsf{raise}~M 
\qquad\qquad	
	M~\mathsf{handle}~(e\,x)\Rightarrow N
\]
The key distinction between coproduct-style and continuation-style exception handling is found in the rule for handling a return value (that is, not an exception), where continuation-style requires that $e$ is not free in $V$, but coproduct-style does not:
\[
	V~\mathsf{handle}~(e\,x)\Rightarrow N ~\rw~ V
\]
That is, continuation-style considers exceptions to be \emph{variables}, where the handler binds $e$ in $V$. Accordingly, the handling construct may be interpreted with a let-binding as follows:
\[
	M~\mathsf{handle}~(e\,x)\Rightarrow N
\quad\mapsto\quad
	\mathsf{let}~e\,x = N~\mathsf{in}~M
\]
By contrast, coproduct-style handling treats exceptions as \emph{constructors} or \emph{constants}. Handling may then be interpreted by a case construct, as follows, where the variable $y$ matches every other case than the exception $e$, returning it unchanged.
\[
	M~\mathsf{handle}~(e\,x)\Rightarrow N
\quad\mapsto\quad
	\mathsf{case}~M~\mathsf{of}~(e\,x\to N~;~y\to y)
\]
Types for the two encodings clarify the difference: for continuation style, in $\mathsf{let}~e\,x = N~\mathsf{in}~M$ we expect $e:A\to B$ for $x:A$, $N:B$, and $M:B$, giving the continuation type $(A\to B)\to B$ to $\lambda e.M$. For coproduct style, in $\mathsf{case}~M~\mathsf{of}~(e\,x~\to~N~;~y~\to~y)$ we have $M:A+B$ for $x:A$ and $N:B$, with $e$ the left injection of the coproduct. The two styles are thus related by the standard higher-order encoding of a coproduct $A+B$ as $(A\to C)\to(B\to C)\to C$.

Exception handling in the literature generally takes one these two forms: interpretations in $\lambda\mu$-calculus~\cite{Ong-Stewart-1997,Bierman-1998,Crolard-1999,VanBakel-2019} are naturally continuation-style, while the exceptions-as-values approach~\cite{Wadler-1985,Spivey-1990,PeytonJones-Reid-Henderson-Hoare-Marlow-1999}, including the exception monad $TX=E+X$, is in coproduct style. The FMC with Choice is in coproduct style, with the following encoding $(-)_\tm$.
\[
	(M~\mathsf{handle}~(e\,x)\Rightarrow N)_\tm \quad=\quad \term{M_\tm;e-><x>.N_\tm}
\]
This extends a standard interpretation of Plotkin's \cbv-$\lambda$-calculus~\cite{Plotkin-1975} into stack calculi~\cite{Douence-Fradet-1998,Power-Thielecke-1999,Heijltjes-2022}, here formulated as $(-)_\val$ and $(-)_\tm$.
\[
\textnormal{Values~}V,W:
\quad
\begin{array}{@{}r@{~}c@{~}l@{}}
               x_\val &=& \term x
\\ (\lambda x.M)_\val &=& \term{<x>.M_\tm}
\end{array}
\qquad
\qquad
\textnormal{Terms~}M,N:
\quad
\begin{array}{@{}r@{~}c@{~}l@{}}
			   V_\tm &=& \term{[V_\val].*}
\\        (M\,N)_\tm &=& \term{M_\tm ; <v>.(N_\tm ; v)} \quad (\term v~\textnormal{fresh})
\end{array}
\]
The interpretation of an application $M\,N$ is to evaluate $\term{M_\tm}$ to a value $\term{[V_\val].*}$; then to pick up $\term{V_\val}$ as $\term v$; evaluate $\term{N_\tm}$ to $\term{[W_\val]}$; and evaluate $\term v$ with $\term{W_\val}$ as its first argument on the stack. The interpretation extends to simple types as follows.
\[
	o_\val~=~\type{1 => 1.*}
\qquad
	(A\to B)_\val ~=~ \type{A_\val => B_\val.*}
\qquad	
	A_\tm ~=~ \type{1 => A_\val.*}
\]
Exceptions $e$ are typed as injections, $\term{e: 1 => 1.e}$, and when applied to a term $\term{M_\tm: 1=>A_\val.*}$ the type becomes $(e\,M)_\tm \rw \term{M_\tm;e : 1 => A_\val.e}$. The \emph{raise} construct is interpreted by the \emph{sum expansion} typing rule: if $\term{M_\tm: 1 => A_\val.e}$ then $\mathsf{raise}~M$ translates to $\term{M_\tm: 1 => A_\val.e + B_\val.*}$ for some type $\type{B_\val}$.

The ML-style reduction rules~\cite[Table 1]{DeGroote-1995} are then simulated in the FMC, preserving types. Four cases are demonstrated below (with the first two slightly adjusted; in the original \emph{raise} alone signifies an exception, but the translation requires an explicit $e$).
\[
\begin{array}{rcl@{\qquad}rcl}
	V\,(\mathsf{raise}\,(e\,W)) &\rw& \mathsf{raise}\,(e\,W)
&	\term{[V_\val].*;<x>.([W_\val].e ; x)} &\rws& \term{[W_\val].e}
\\
	(\mathsf{raise}\,(e\,V)) W &\rw& \mathsf{raise}\,(e\,V)
&	\term{[V_\val].e ; <x>.([W_\val].* ; x)} &\rws& \term{[V_\val].e}
\\
	V~\mathsf{handle}~(e\,x)\Rightarrow N &\rw& V
&	\term{[V_\val].*~;~e-><x>.N_\tm} &\rw& \term{[V_\val].*}
\\
	\mathsf{raise}~e\,V~\mathsf{handle}~(e\,x)\Rightarrow N &\rw& \{V/x\}N
& \term{[V_\val].e~;~e-><x>.N_\tm} & \rws &\term{\{V_\val/x\}N_\tm}	
\end{array}
\]

The relation with \emph{algebraic effects}~\cite{Plotkin-Power-2001} and \emph{effect handlers}~\cite{Plotkin-Pretnar-2013} is not yet fully characterised. It is clear that effect handlers should be considered as continuation-style, and a handler understood as a binding construct for effect operations. A tentative interpretation is the following, where operators $\mathsf{op}$ are variables and a handler $H$ is a \emph{prefix} of push- and pop-actions, binding in $N$.
\[
\begin{array}{r@{~=~}ll}
		          \mathsf{op}\,V\,(x.M) & \term{[V].{\mathsf{op}}~;~<x>.M}
\\  N~\mathsf{handle}~H~\mathsf{to}~x.M & \term{H.N\,;\,<x>.M}
\\         (\mathsf{op}\,x\,k\to N) + H & \term{[<x>.\{*/k\}N].<{\mathsf{op}}>.H}
\end{array}
\]
Note that the continuation $\term k$ in $\term N$ is discarded, and replaced with \emph{skip}, and correspondingly the continuation $K$ in $\mathsf{op}\,V\,K$ is composed sequentially, rather than passed as an argument. This interprets handlers directly in their \emph{generic} form~\cite{Plotkin-Power-2003,Plotkin-Pretnar-2013}. The interpretation is correct for handlers satisfying the requisite law.
\[
	\mathsf{op}~V~(x.M)~;~K \quad=\quad \mathsf{op}~V~(x.(M;K))
\]
\emph{Shallow handlers}~\cite{Hillerstrom-Lindley-2018} superficially appear closer to coproduct-style, since handlers become \emph{affine} (used at most once). However, for subtle reasons, reductions are not fully compatible with the expected interpretation in \emph{choice} constructs. This would be as below, where an effect operator is a \emph{jump} and a handler $H$ is a postfix sequence of \emph{joins}. The discrepancy needs further work to fully understand.
\[
\begin{array}{r@{~=~}ll}
		                 \mathsf{op}\,V & \term{[V].{\mathsf{op}}}
\\                  N~\mathsf{handle}~H & \term{N\,;\,H}
\\         H + (\mathsf{op}\,x\,k\to N) & \term{H\,;\,{\mathsf{op}} -> <x>.\{*/k\}N}
\\   			 \mathsf{return}~x\to N & \term{* -> <x>.N}
\end{array}
\]

\section{Conclusion}

This note has given an overview of current progress on the \emph{choice} extension for the Functional Machine Calculus. At present the design is complete, proofs are expected, and a start has been made on a close comparison with previous work. These are promising: the calculus is expected to fulfil the six aims given in the introduction (\emph{confluence}, \emph{types}, \emph{minimality}, \emph{operational semantics}, \emph{seamless integration}, and \emph{first-order restriction}), and the discussion in the previous section is starting to give an interesting and revealing perspective on past work. Clearly, this is only a starting point, and there is much more to investigate.

\subsection*{Acknowledgements}

The author is greatly indebted to Nicolas Wu for explaining effect handlers.


\bibliographystyle{./entics}
\bibliography{FMC}

\end{document}